\documentclass[aps,superscriptaddress,prc,twocolumn,nofootinbib]{revtex4-2}

\usepackage[breaklinks,colorlinks,urlcolor=blue,citecolor=blue,linkcolor=blue]{hyperref}
\usepackage{amsmath,bm}
\usepackage{graphicx}
\usepackage{epstopdf}
\usepackage{subfigure}
\usepackage{epsfig}
\usepackage{amsmath,amssymb,amsfonts}
\usepackage{color}
\usepackage[utf8]{inputenc}
\usepackage{verbatim}
\usepackage{array}
\usepackage{multirow}
\begin{document}

\renewcommand{\topfraction}{0.85}
\renewcommand{\textfraction}{0.1}
\renewcommand{\floatpagefraction}{0.75}

\title{The Fox-Wolfram Moment of Jet Production in Relativistic Heavy Ion Collisions}

\date{\today  \hspace{1ex}}
\author{Wei-Xi Kong}
\affiliation{
  Key Laboratory of Quark \& Lepton Physics (MOE) and Institute of Particle Physics,
  Central China Normal University, Wuhan 430079, China
}

\author{Ben-Wei Zhang}
\email{bwzhang@mail.ccnu.edu.cn}
\affiliation{
  Key Laboratory of Quark \& Lepton Physics (MOE) and Institute of Particle Physics,
 Central China Normal University, Wuhan 430079, China
}

\begin{abstract}
Fox-Wolfram Moments (FWMs) are a set of event shape observables that characterize the angular distribution of energy flow in high-energy collisions. In this work, we present the first theoretical investigation of FWMs for multi-jet production in relativistic heavy ion collisions. In this work, jet productions in p+p collisions are computed with a Monte Carlo event generator SHERPA, while the Linear Boltzmann Transport model is utilized to simulate the multiple scattering of energetic partons  in the hot and dense QCD matter. The event-normalized distributions of the lower-order FWM, $H_1^T$ in p+p and Pb+Pb collisions are calculated.  It is found that for events with jet number $n_\text{jet} = 2$ the $H_1^T$ distribution in Pb+Pb is suppressed at small $H_1^T$ while enhanced at large $H_1^T$ region as compared to p+p.  For events with $n_\text{jet}>2$, the jet number reduction effect due to jet quenching in the QGP decreases the $H_1^T$ distribution at large $H_1^T$ in Pb+Pb relative to p+p. The medium modification of the Fox-Wolfram moment $H_1^T$ for events with $n_\text{jet}\ge 2$ are also presented, which resemble those of events with $n_\text{jet} = 2$. Its reason is revealed through the relative contribution fractions of  events with different final-state jet numbers to $H_1^T$.
\end{abstract}

\pacs{13.87.-a; 12.38.Mh; 25.75.-q}

\maketitle

\section{introduction}
\label{sec:Intro}

The quark-gluon plasma (QGP), a new state of QCD matter composed of de-confined quarks and gluons, has been a principal focus of contemporary research in high-energy nuclear physics for  the last thirty years~\cite{Gyulassy:2004zy, Muller:2012zq, Wang:2016opj, Busza:2018rrf}. Highly energetic partons generated in the initial collision traverse the QGP medium, experiencing energy loss through interactions with constituent partons of the medium. This phenomenon, recognized as jet quenching, serves as a potent investigative tool for probing the characteristics of the QGP~\cite{Bjorken:1982tu, Gyulassy:1990ye, Wang:1991xy, Baier:1996sk, Gyulassy:2003mc, Qin:2015srf, Wicks:2005gt, Bass:2008rv, Cao:2020wlm, Baier:2001yt, JETSCAPE:2020ttu, Pablos:2022piv, Luo:2023nsi, Xie:2022ght, Xie:2024xbn, Zhang:2021xib}. Various observables have emerged to quantitatively investigate the phenomena of jet quenching. 
Initially, investigations focused on how the medium modifies hadron spectra and hadron correlations, such as the suppression of leading hadron spectra, modifications of di-hadron correlations, and other hadronic observables studies~\cite{Brenner:1981kf, STAR:2002svs, Albacete:2010bs, ALICE:2010yje, ALICE:2011gpa, STAR:2010mfd, Xing:2020hwh, Dai:2015dxa, Xing:2024qcr, Li:2024wqq, Zhang:2022rby}. With the sustained running of the Large Hadron Collider (LHC) and the development of theory, various observables at jet level have been developed. These observables respectively concentrate on the production or correlation of inclusive jets~\cite{ATLAS:2012tjt, ALICE:2015mdb, STAR:2014wox, STAR:2007rjc, Kang:2016mcy, STAR:2017hhs, CMS:2016uxf, ALICE:2019qyj, ATLAS:2013ssy, Dasgupta:2016bnd, He:2018xjv, JETSCAPE:2022jer}, dijets~\cite{Magestro:2005vm, ATLAS:2010isq, He:2011pd, Milhano:2015mng, Dai:2018mhw, Li:2024uzk}, and tagged jets~\cite{Wang:1996yh, PHENIX:2009cvn, STAR:2009ojv, Neufeld:2010fj, Dai:2012am, CMS:2017eqd, Zhang:2018urd, Wang:2019xey, Li:2022tcr, ALICE:2021aqk, Dai:2022sjk, Wang:2021jgm, Yang:2021qtl, Xiao:2024ffk, Liu:2021dpm, Gao:2023lhs, JETSCAPE:2024rma}, with some also scrutinizing the structure and substructure of aforementioned categories of jet events~\cite{ATLAS:2013uet, CMS:2013lhm, Cal:2019hjc, CMS:2012nro, ATLAS:2014dtd, ATLAS:2018bvp, Kang:2016ehg, Larkoski:2017jix, Kogler:2018hem, Milhano:2017nzm, Feige:2012vc, Li:2017wwc, Wang:2022yrp, Kang:2023ycg, Yang:2023dwc, Yang:2022nei, Yang:2022yfr, JETSCAPE:2023hqn}.

In addition to these typical experimental observables under sustained attention, there exists a category of measurements capturing the geometric shape of the entire jet event, known as jet event shape observables~\cite{Banfi:2010xy, CMS:2011usu, ATLAS:2012tch, CMS:2013lua}. The early studies of event shape observables were primarily conducted in electron-positron collisions, deep inelastic scattering (DIS) and proton-antiproton collisions~\cite{Catani:1992ua, OPAL:2004wof, Bauer:2008dt, Ellis:1980nc, Dasgupta:2002dc, H1:1999wfh, DELPHI:2003yqh, DELPHI:1996oqw, CDF:1991etg, Catani:1992jc}. With extensive research in high-energy nuclear physics ongoing at the LHC, some theoretical  explorations involving jet event shape observables in heavy ion collisions have emerged in recent years~\cite{Chen:2020pfa, Mallick:2020dzv, Mallick:2020ium, Prasad:2021bdq, Kang:2023qxb}.  These studies hold significant importance in understanding the properties of nuclear medium~\cite{Qiu:2011iv, Berger:2003iw, ALICE:2020iug, ALICE:2015lib, ALICE:2018gif} and elucidating the dynamics of extreme relativistic collisions~\cite{Schukraft:2012ah, Jia:2014jca}. 

In this work we extend the studying of medium modifications of jet shapes in heavy-ion collisions to Fox-Wolfram moments (FWMs), a set of rather unique event shape measurements characterized by expansions in terms of spherical harmonics~\cite{Fox:1978vu}.  The application of FWMs has been extensive in collider physics, particularly in high-energy physics domains, including studies in Higgs physics~\cite{Bernaciak:2012nh, Bernaciak:2013dwa}, top physics~\cite{Field:1996nq}, and general investigations in $e^+e^-$ collisions~\cite{Nagy:1997mf, Li:2020vav} as well as hadron collisions~\cite{Spiller:2015axa}.
We make the first theoretical attempt to calculate the medium modifications of the event-normalized distribution of the lower-order FWM, $\text{FWM}_\text{1st}^\text{T}$ (denoted as $H_1^T$ in the subsequent text) for  events with $2$ jets and with $>$ 2 jets, by comparing $H_1^T$ distributions in p+p and Pb+Pb collisions at $\sqrt{s}=5.02$ TeV. For events with $2$ jets, the results reveal a suppression in the small $H_1^T$ region and an enhancement in the large $H_1^T$ region for the distribution in Pb+Pb collisions compared to p+p collisions. Meanwhile, for events with $>$ 2 jets, there is a suppression of the Pb+Pb collisions relative to the p+p collisions in the large $H_1^T$ region. Furthermore, when the jet multiplicity is unspecified, i.e., events with $\ge$ 2 jets, the modifications resemble that of events with 2 jets. And computations regarding the relative contributions of the two types of events with respect to $H_1^T$ elucidate the underlying reasons.

Our study on $H_1^T$ serves as a valuable complement to other event shape observables in heavy-ion collisions such as the jet broadening $B_{tot}$ and the transverse sphericity $S_{\perp}$ which have been investigated recently~\cite{Kang:2023qxb, Chen:2020pfa}. The advantage of this work lies in the fact that, compared to the jet broadening $B_{tot}$ which gives no zero results only for events with the jet number $n_\text{jet} \geq 3$~\cite{Kang:2023qxb}, $H_1^T$ covers the events both with the jet number $n_\text{jet} = 2$ and $n_\text{jet} \geq 3$ . In addition, compared to the transverse sphericity, which characterizes the similar pattern of energy flow~\cite{Chen:2020pfa} as the FWM, $H_1^T$ may exhibit a larger magnitude of medium modification.

The subsequent organization of the content is as follows. In the Sec.~\ref{sec:FWM}, we introduce FWMs, from which we extract a specific observable of our focus. And we discuss and elucidate its physical interpretation in p+p collisions. In Sec.~\ref{sec:results}, the results and discussions in p+p and Pb+Pb collisions are presented, followed by the summary in Sec.~\ref{sec:sum}.

\section{FWM\lowercase{s} analysis and $\text{FWM}_\text{1st}^\text{T}$ in \lowercase{p+p} collisions}
\label{sec:FWM}

Giving by a superposition of spherical harmonics $Y_\ell^m(\Psi_i)$, FWMs is a set of such observables~\cite{Fox:1978vu}

\begin{alignat}{1}
	H_\ell^o = \frac{4\pi}{2\ell+1}
	\sum_{m=-\ell}^\ell \;
	\left| \sum_{i=1}^{n_\text{jet}}  W_i^o \; Y_\ell^m(\Psi_i) 
	\right|^2 \;
	\label{eq:fwm_def1}
\end{alignat}

FWMs are sensitive to the number of jets, angular correlation of jets, and energy distribution of jets. They systematically describe geometric correlations in terms of spherical harmonics. The inner sum is over the final state objects (here are jets). There are various schemes for the construction of weight factors $W_{i}^o$ as shown in the literature Ref.~\cite{Bernaciak:2012nh, Spiller:2015axa}. The `o' here serves as a marker for a different scheme. It represents the individual components of momentum or the vector sum of different components, such as $p_x$, $p_y$, $p_z$, $p_T$, and $|\vec{p}|$. The typical schemes are as follows:

\begin{alignat}{2}
  W_{i}^o
  = \frac{p^o_i}{\sum_{k=1}^{n_\text{jet}} p^o_k}
  \label{eq:WT}
\end{alignat}

Since the angular variable $\Psi_i$ depends on the choice of axis, we reformulate Eq.~\eqref{eq:fwm_def1} to achieve an axis-independent formulation by utilizing the inherent relationship between spherical harmonics and Legendre polynomials. Specifically, the sum over spherical harmonics can be rewritten in terms of Legendre polynomials as  

\begin{equation}
\sum_{m=-\ell}^{\ell} Y^m_\ell (\Psi_i) Y^{m*}_\ell (\Psi_j) = \frac{2\ell+1}{4\pi} P_\ell(\cos\Delta\Psi_{ij}).
\end{equation}

By substituting this relation into Eq.~\eqref{eq:fwm_def1}, we obtain the following simplified expression for the FWMs:

\begin{alignat}{3}
	H^o_\ell = \sum_{i,j=1}^{n_\text{jet}} \; W_{ij}^o \; P_\ell(\cos \Delta\Psi_{ij}) 
    \text{,} \quad 
	W_{ij}^o = \frac{p^o_i p^o_j}{(\sum_{k=1}^{n_\text{jet}} p^o_k)^2}
	\label{eq:fwm_def2}
\end{alignat}

This formulation expresses FWMs as a weighted sum over Legendre polynomials, making it more convenient for analyzing angular correlations in jet events while remaining independent of the reference axis.

According to different weight construction schemes, we can choose different angle schemes $\Delta\Psi_{ij}$ to measure the distance between jets. Here, $\Delta\Psi_{ij}$ can take on values of $\Delta\phi_{ij}$, $\Delta\theta_{ij}$, and $\Delta\Omega_{ij}$, which adhere to the spherical coordinates equation $\cos\Delta\Omega_{ij} = \cos\theta_i\cos\theta_j + \sin\theta_i\sin\theta_j\cos (\phi_i-\phi_j)$. The angle $\phi_i$ and $\theta_i$ represent the azimuth angle of jet and angle between jet and beam direction, among which the latter satisfies the equation with jet rapidity $\eta_i = -\ln(\tan\frac{\theta_i}{2})$. So, it is straightforward to ascertain that $\Delta\phi_{ij}$ and $\Delta\theta_{ij}$ represent the azimuthal separation and the difference in rapidity between jet i and j. And $\Delta\Omega_{ij}$ represents the distance between them in three-dimensional momentum space.

Given our emphasis on jet physics, the transverse momentum and azimuthal separation $\Delta\phi_{ij}$ of jets are employed in the construction of the FWMs. The values of $i,j (1, 2, 3, ...)$ represent jets ordered by descending transverse momentum. And the marker `o' changes to the `T':

\begin{equation}
\begin{aligned}
  H_\ell^{T} & = \sum_{i,j=1}^{n_\text{jet}} \; W_{ij}^T \; P_{\ell}(\cos \Delta\phi_{ij}) 
  \\ & = \sum_{i,j=1}^{n_\text{jet}} \; \frac{p_{T i}p_{T j}}{\left(\sum_{k=1}^{n_\text{jet}} p_{T k}\right)^2} \; P_{\ell}(\cos \Delta\phi_{ij})
  \label{eq:fwm_def3}
\end{aligned}
\end{equation}

\begin{figure*}
  \centering
  \hbox{
	\includegraphics[scale=0.22]{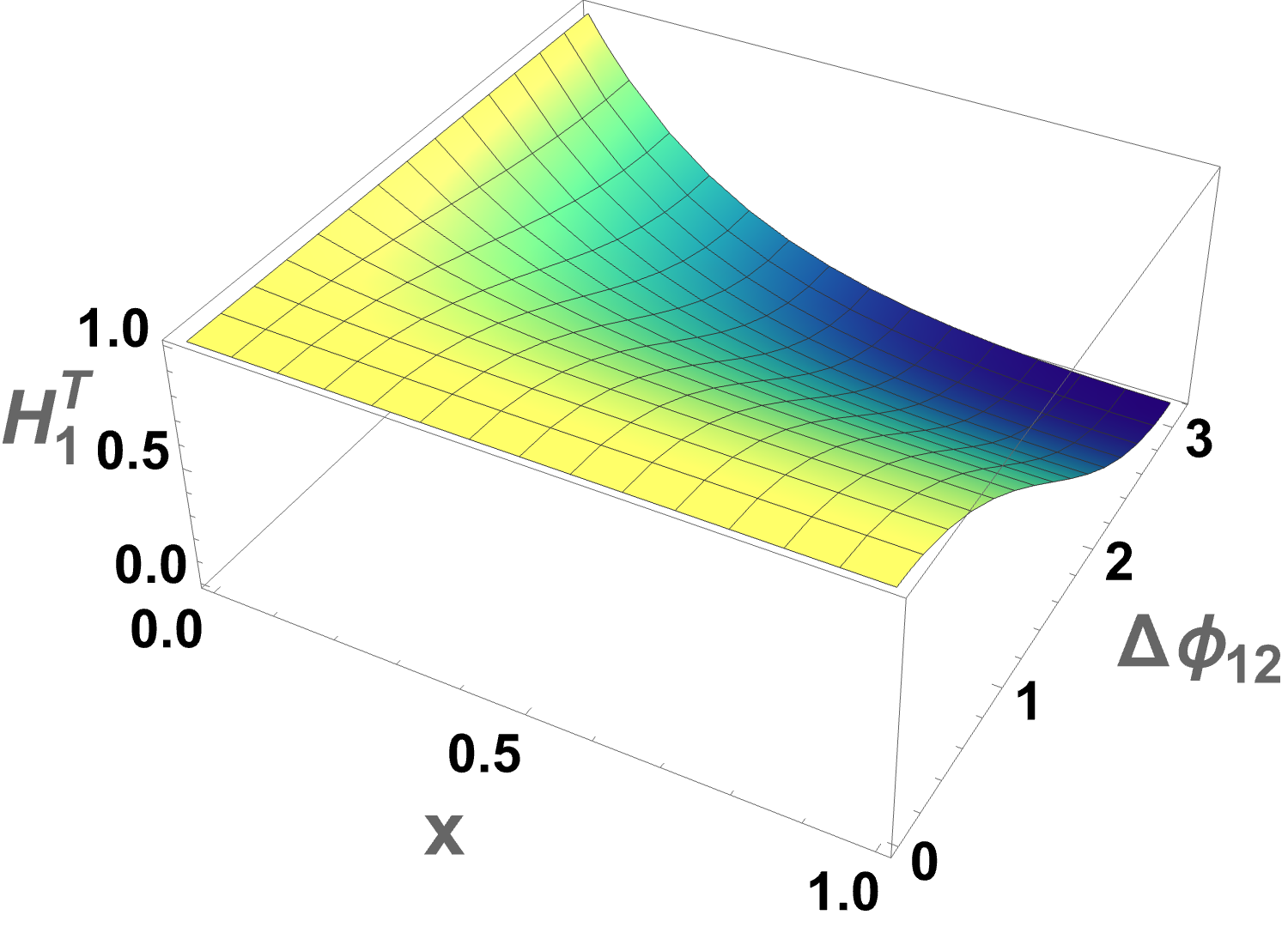}
	\includegraphics[scale=0.22]{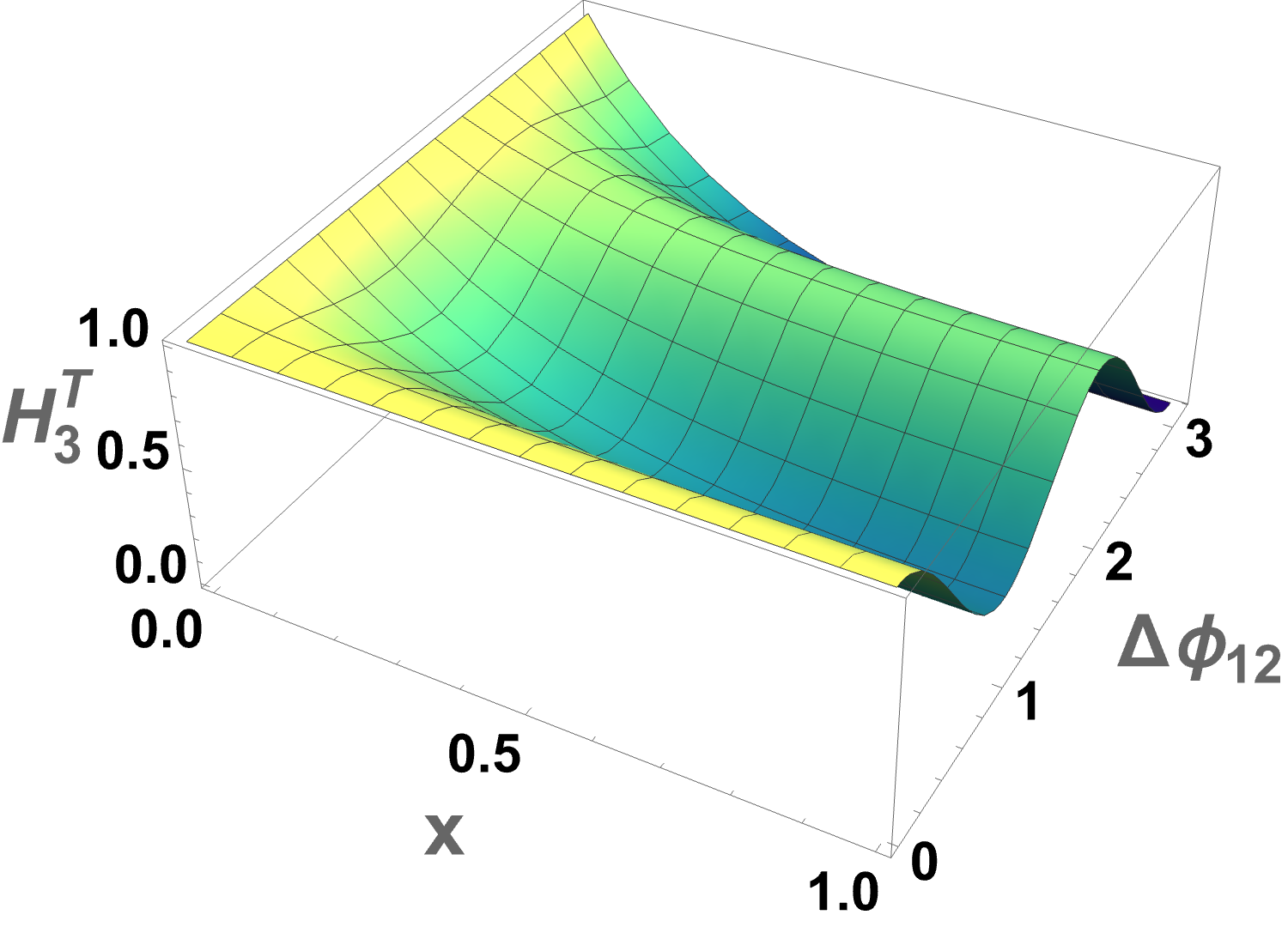}
	\includegraphics[scale=0.22]{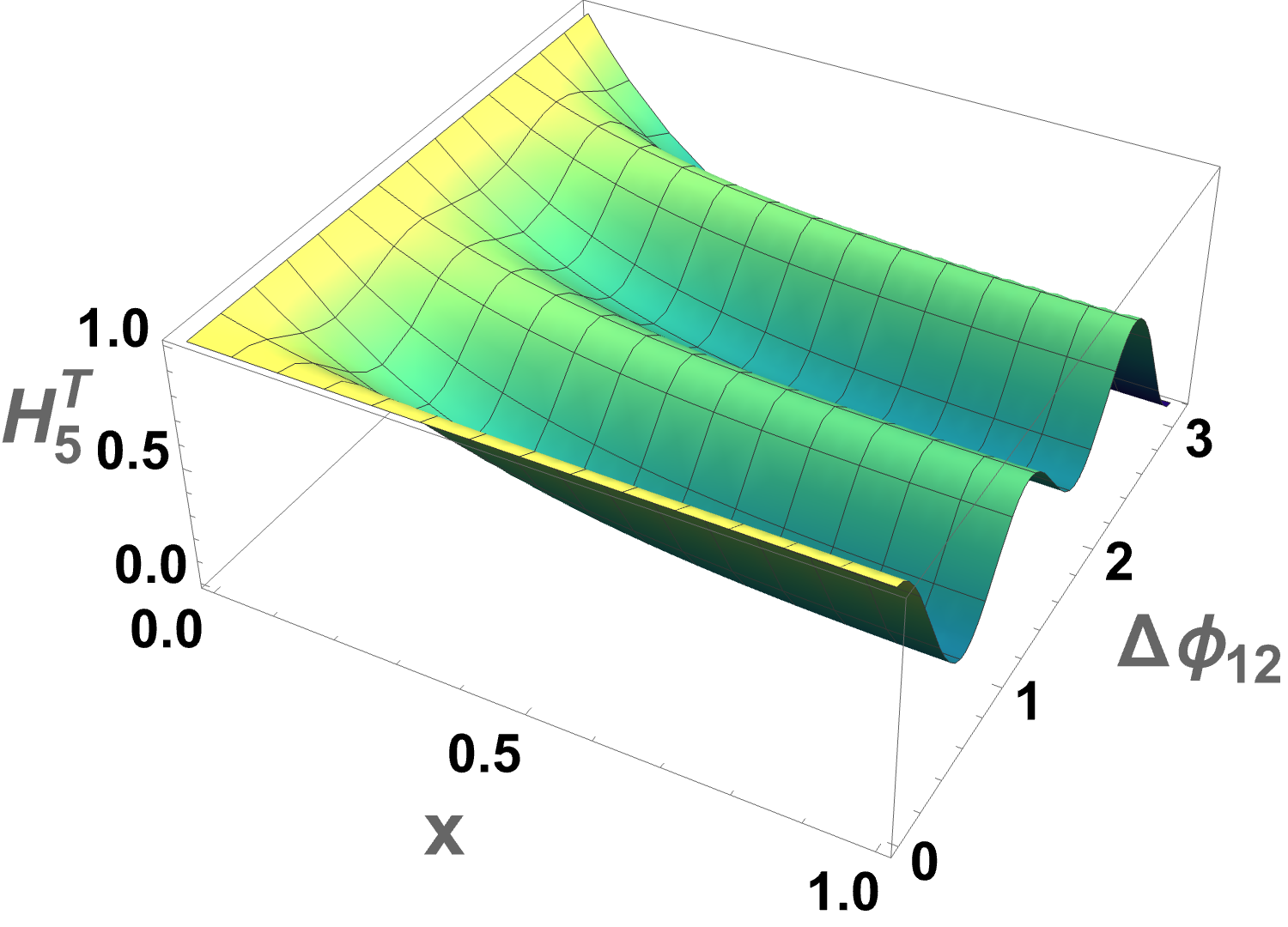}
  }
  \hbox{
	\includegraphics[scale=0.22]{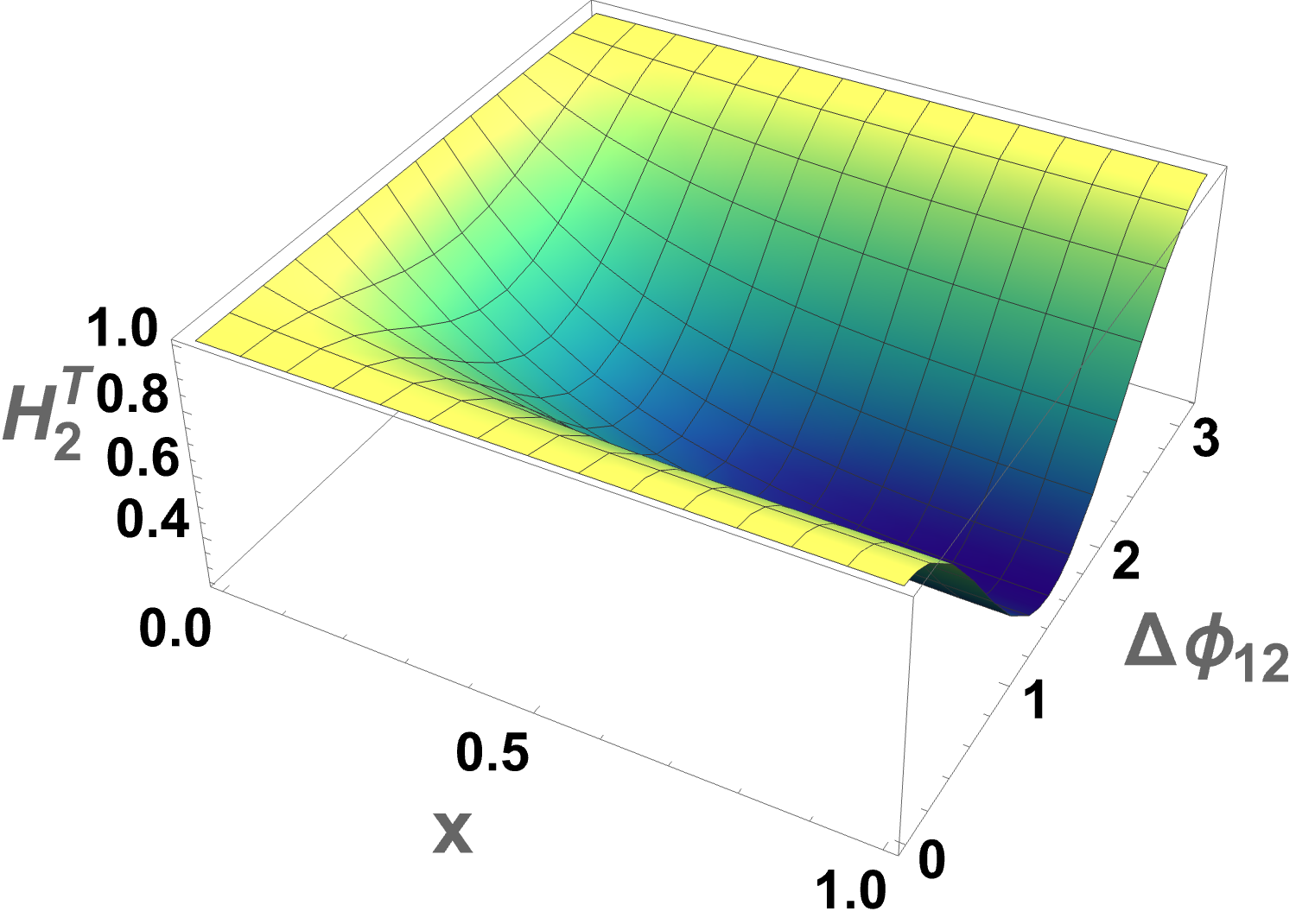}
	\includegraphics[scale=0.22]{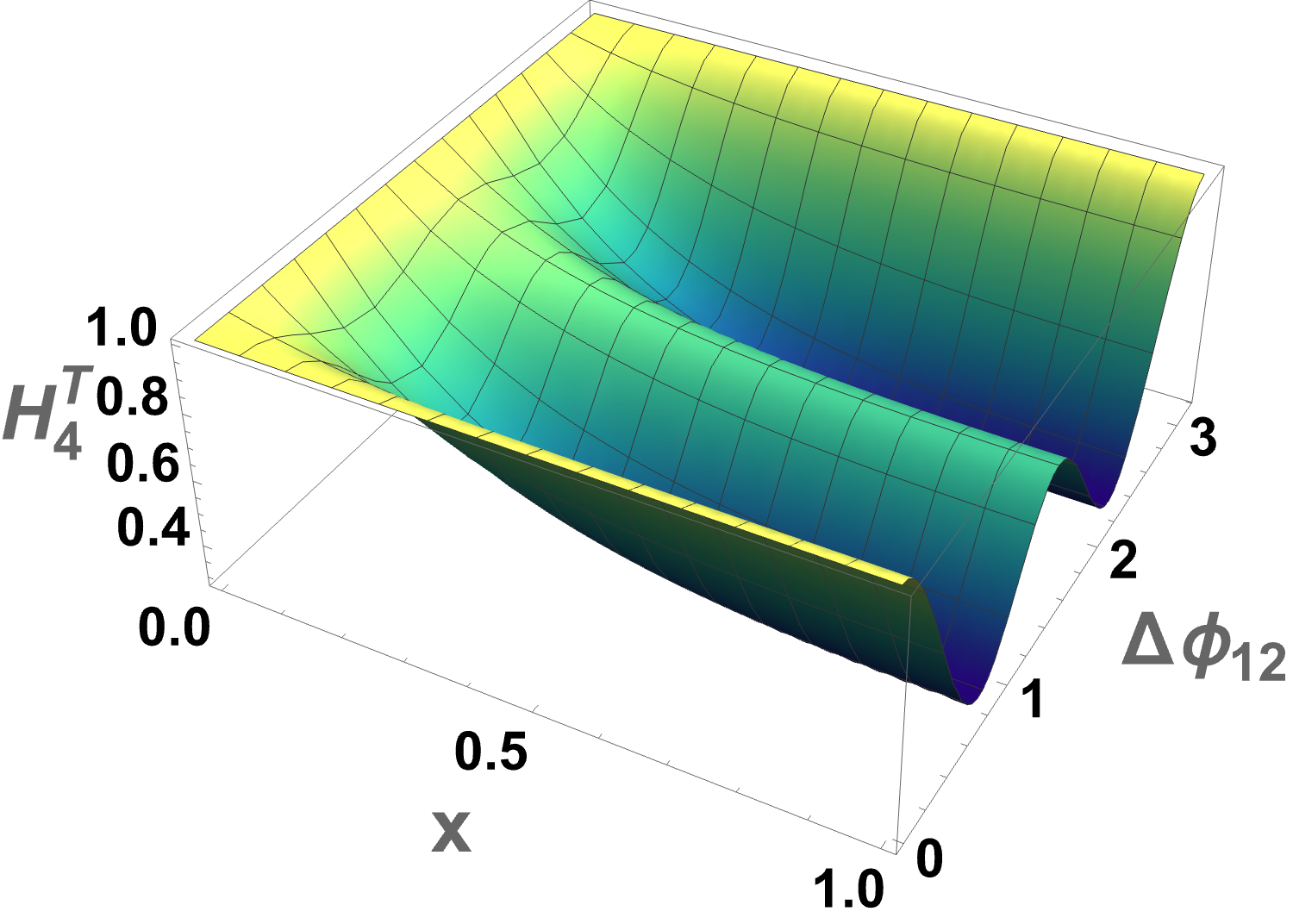}
	\includegraphics[scale=0.22]{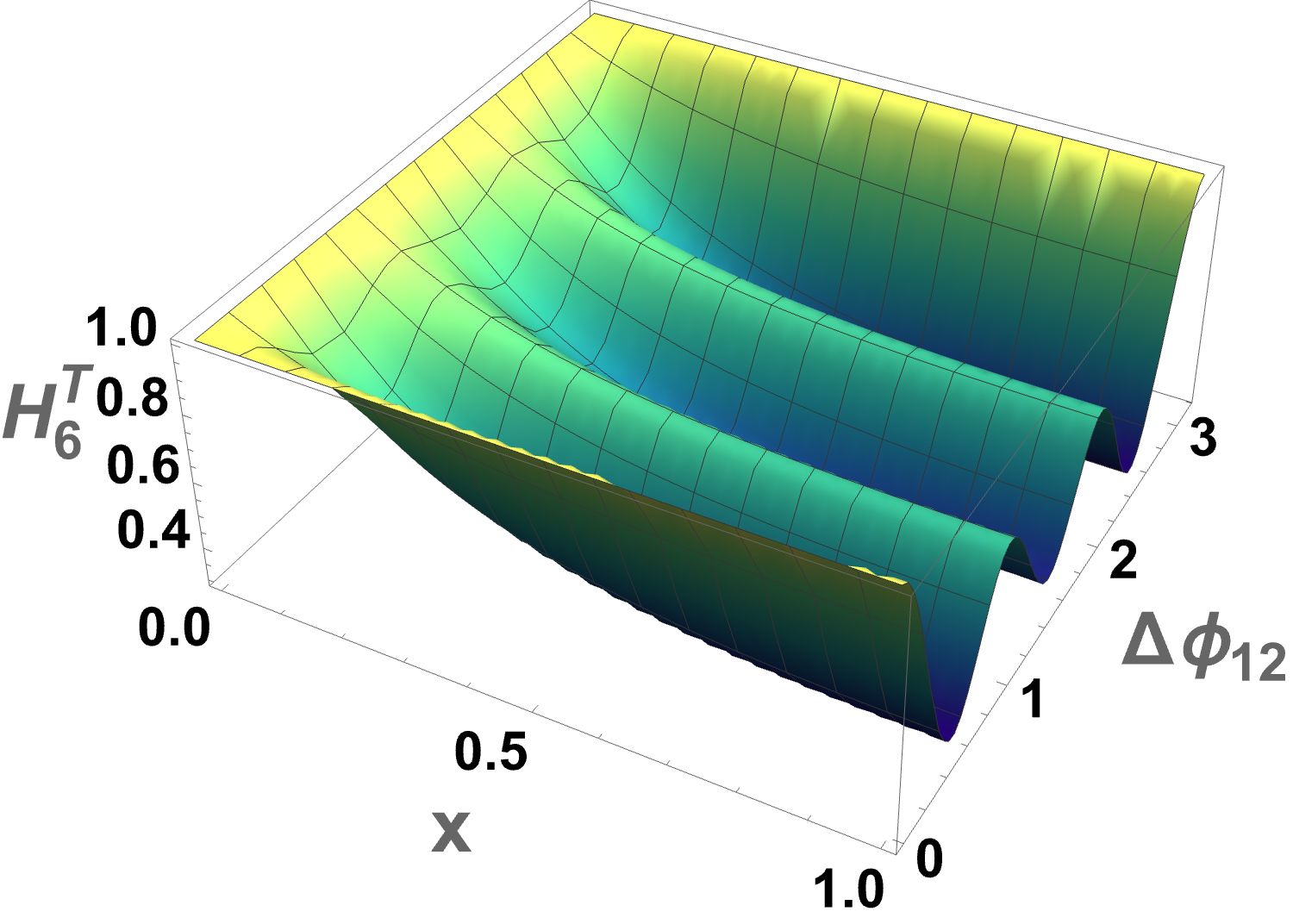} 
  }
	\caption{The analytic result of the FWMs $H_\ell^T$ ($n_\text{jet}$ = 2) for $\ell = 1,2,3,4,5,6$.}
	\label{fig:test}
\end{figure*}

Let's first analyze the toy model given by Eq.\eqref{eq:fwm_def3} when $n_\text{jet}$ is equal to 2 as shown in FIG.~\ref{fig:test}. For simplicity, we denote $p_{T 2} = x\cdot p_{T 1} $ which simplify the functional form of the weights $W_{ij}^T$. Expanding the sum in the definition Eq.\eqref{eq:fwm_def3} with weights expressed by transverse momentum yields

\begin{alignat}{5}
	H_\ell^{T}(n_\text{jet} = 2) = \frac{1+2 x P_\ell(\cos\Delta\phi_{12})+x^2}{1+2x+x^2}
	\label{eq:fwm_toy}
\end{alignat}

If $\ell = 0$, then the Legendre polynomial $P_0 = 1$, resulting in $H_0^T = 1$, which is not meaningful for our study. Due to the properties of the Legendre polynomials $P_{\ell}$, the FWMs $H_{\ell}^T$ exhibit strongly oscillatory behavior, as shown in FIG.~\ref{fig:test}. The larger value of $\ell$, the stronger oscillatory. This can be viewed as a change in the resolution with which the FWMs probe the structure of the QCD jet geometry~\cite{Bernaciak:2012nh}. For this reason, the study of FWMs of different orders was used to provide sufficient resolution to discriminate signal and backgrounds in Higgs and t$\bar{\mathrm{t}}$ production researches~\cite{Bernaciak:2012nh,Field:1996nq}. 

As $\Delta \phi_{12}$ approaches $\pi$, $H_{\ell}^T$ approaches 1 (even moments) or 0 (odd moments). And all weights $W_{ij}^T$ ensure that the range for the moments is $0 \le H_\ell^T \le 1$ and preserve the different shapes of $H_\ell^T$ for even or odd moments. As pioneers in the study of this observable in the field of jet quenching, we first focus on the monotonic relationship between the lower-order $H_1^T$ and the azimuthal angle distance to investigate the jet event shape. The definition is as follows:

\begin{equation}
  \begin{aligned}
  H_1^{T} & = \sum_{i,j=1}^{n_\text{jet}} \; \frac{p_{T i}p_{T j}}{\left(\sum_{k=1}^{n_\text{jet}} p_{T k}\right)^2} \; P_{1}(\cos \Delta\phi_{ij})\\
  & = \sum_{i,j=1}^{n_\text{jet}} \; \frac{p_{T i}p_{T j}}{\left(\sum_{k=1}^{n_\text{jet}} p_{T k}\right)^2} \; \cos \Delta\phi_{ij}
  \label{eq:fwm_def4}
  \end{aligned}
\end{equation}

\begin{figure*}
  \hbox{
      \includegraphics[scale=0.23]{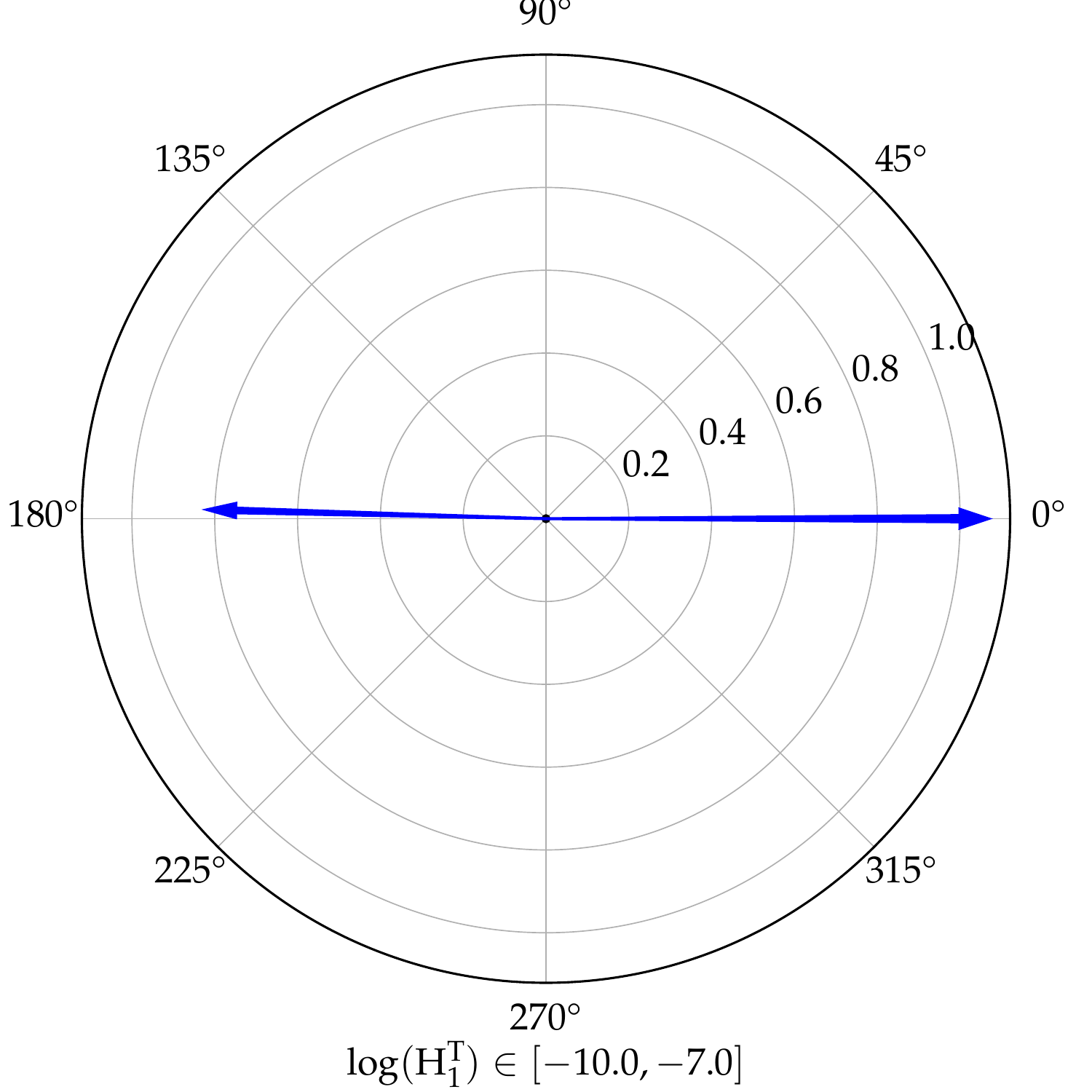}
      \includegraphics[scale=0.23]{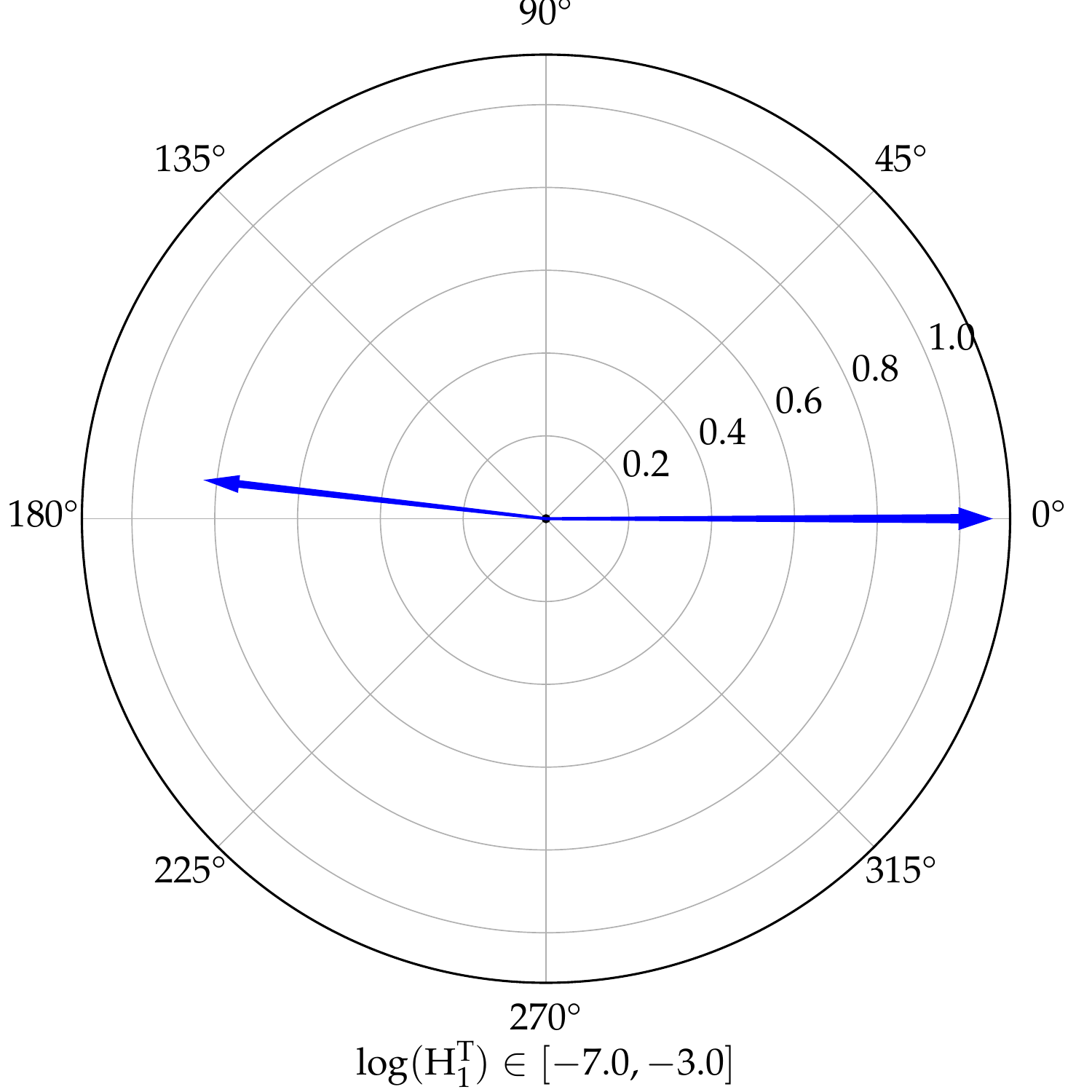}
      \includegraphics[scale=0.23]{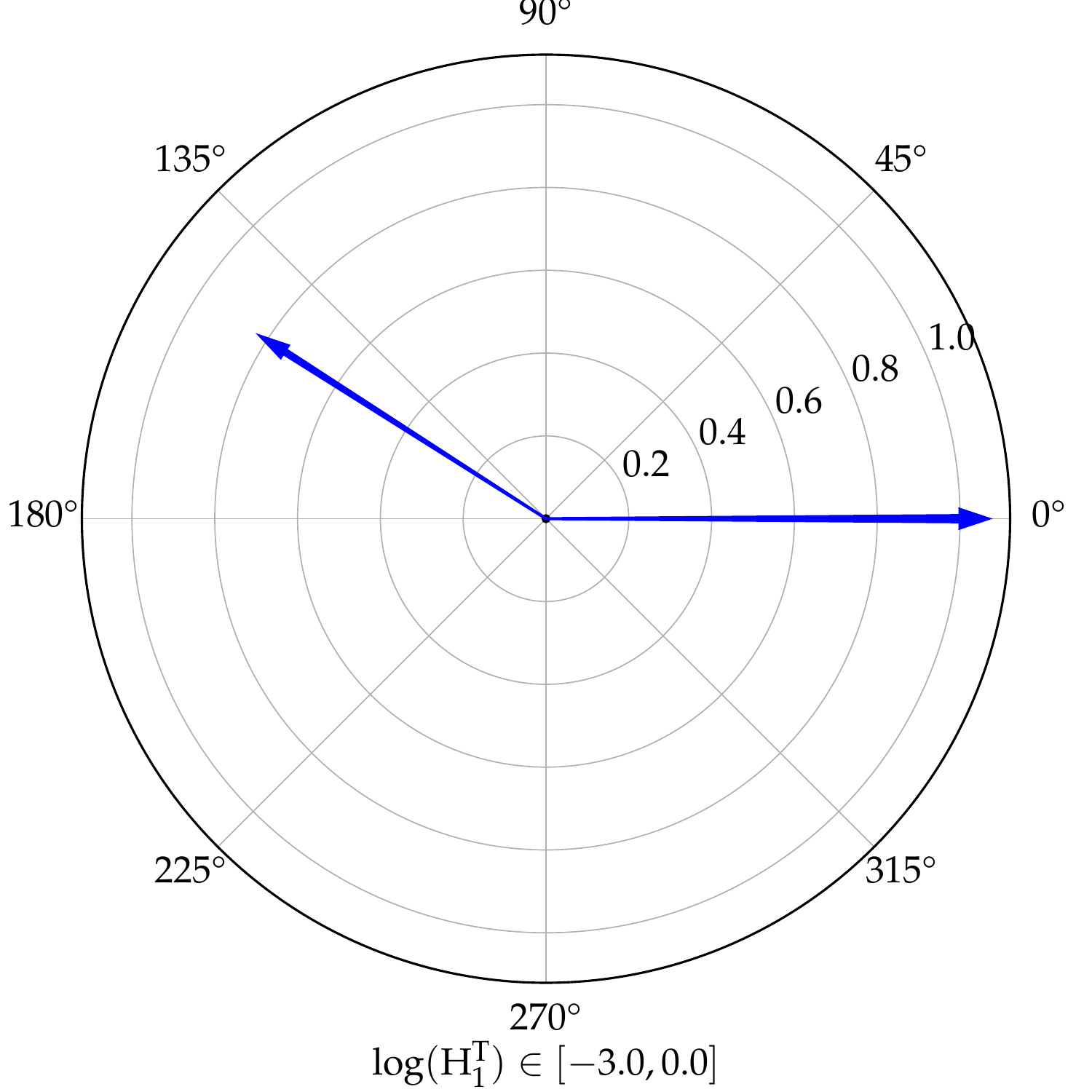}
  }
  \hbox{
      \includegraphics[scale=0.23]{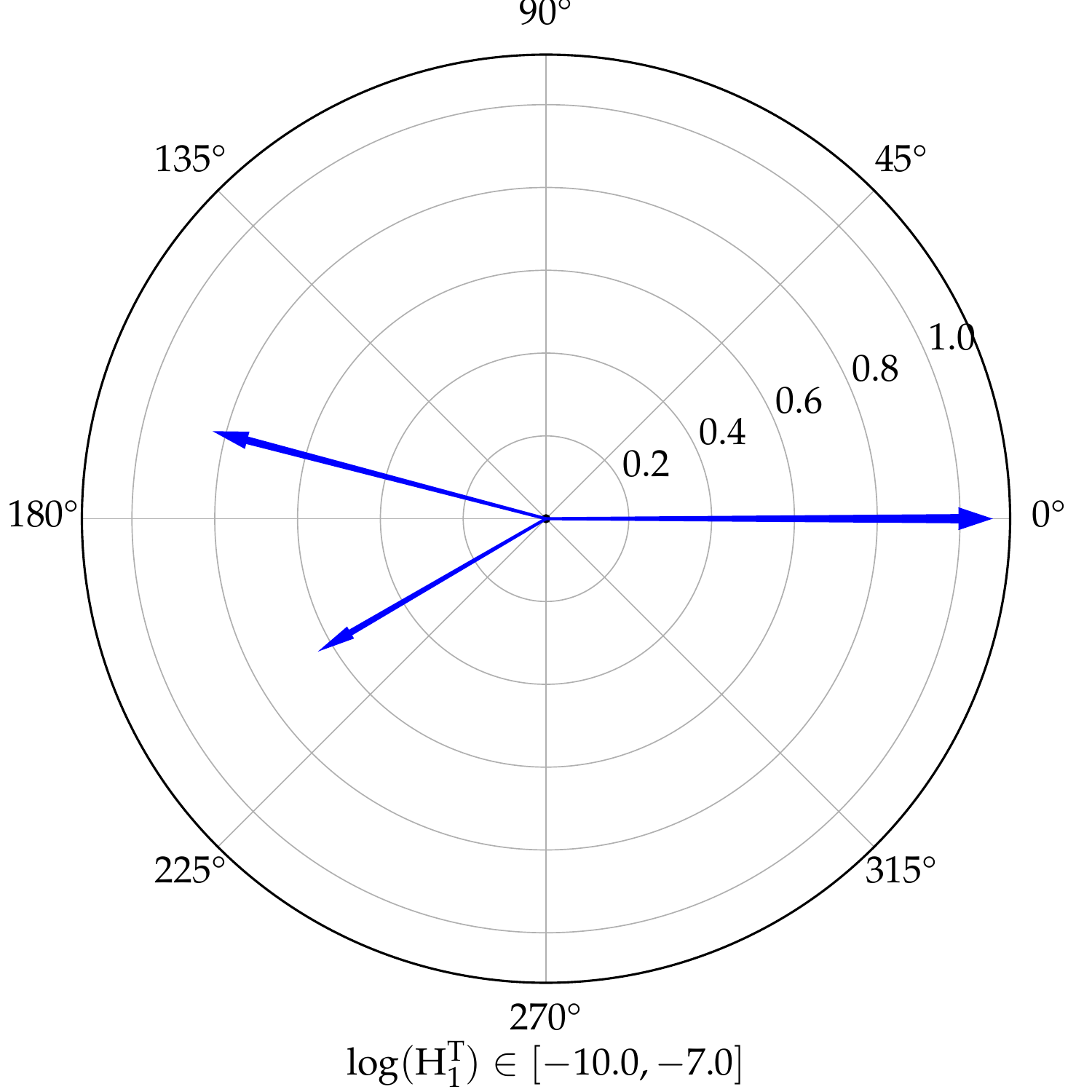}
      \includegraphics[scale=0.23]{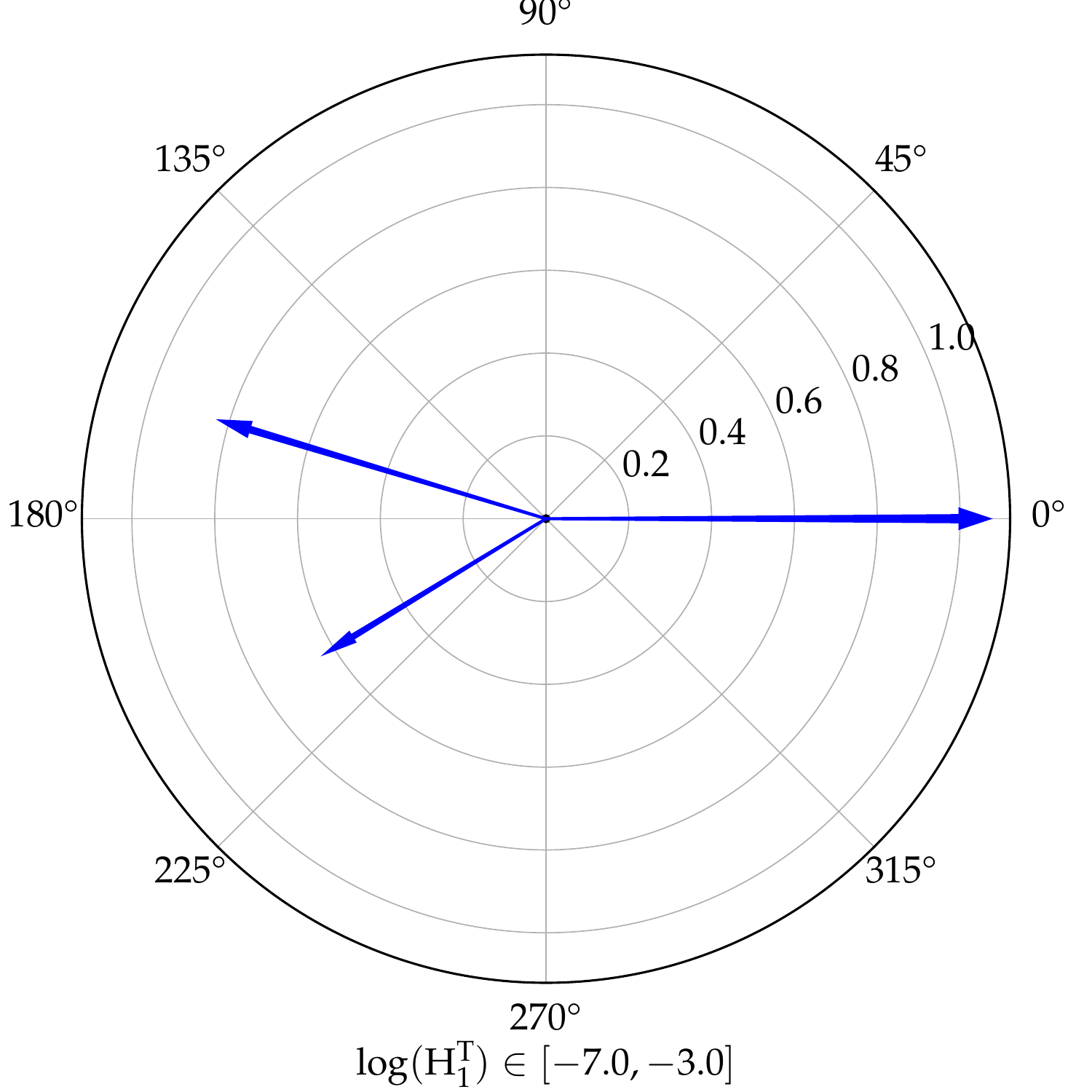}
      \includegraphics[scale=0.23]{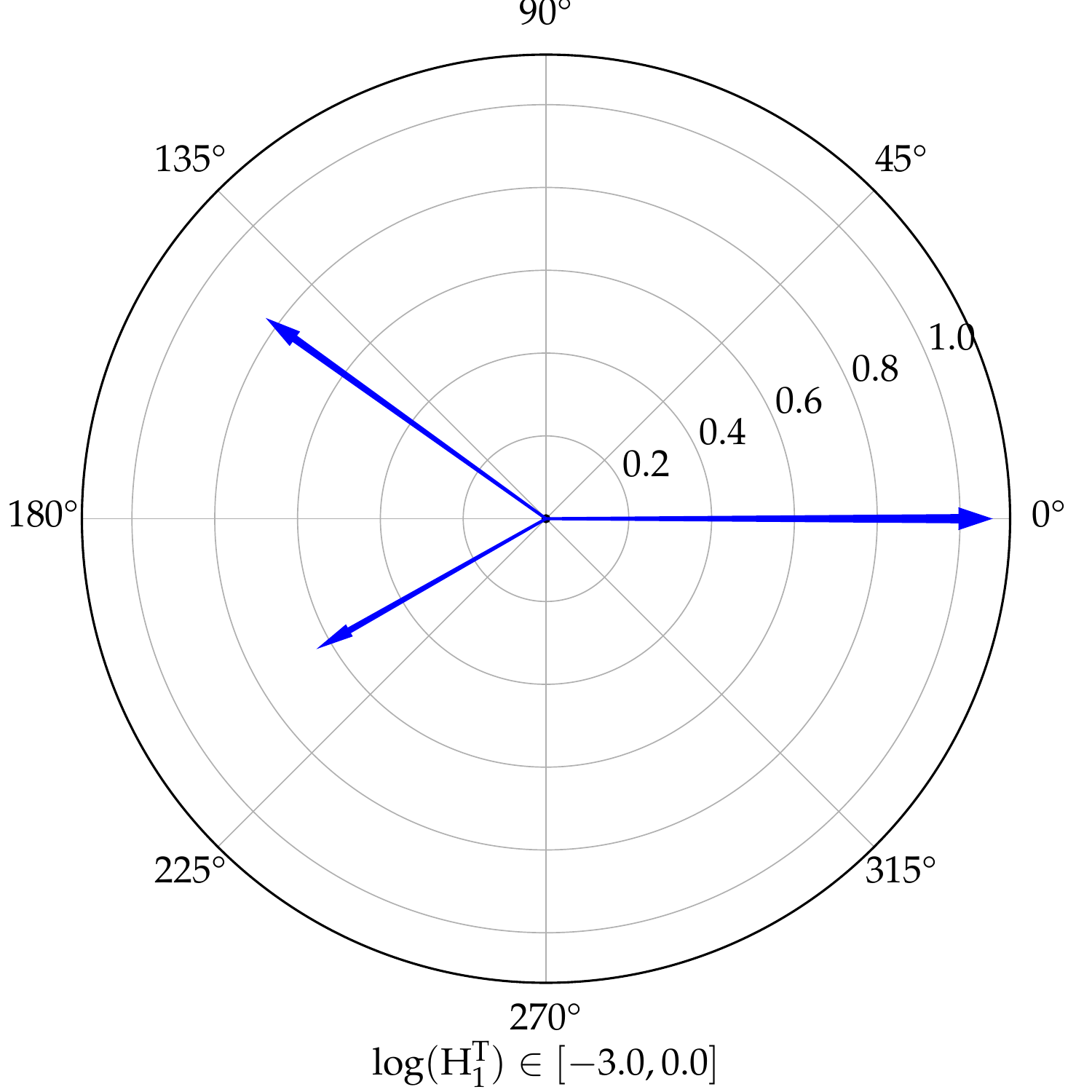}     
  }
  \caption{In the transverse plane, schematic diagrams illustrating the spatial configurations of $n_\text{jet} = 2$ and $n_\text{jet} = 3$ events with polar coordinates across three consecutive $H_1^T$ intervals.}
  \label{config:jet}
\end{figure*}

The definition of \( H_1^T \) differs for events with different \( n_\text{jet} \). It is straightforward to see that for \( n_\text{jet} = 2 \) events, the definition of \( H_1^T \) is obtained by replacing \( P_\ell(\cos\Delta\phi_{12}) \) in Eq.~\eqref{eq:fwm_toy} with \( \cos\Delta\phi_{12} \). However, due to the increase in the number of variables, the definition of events \( n_\text{jet} = 3 \) is more complex, which will be discussed in Appendix~\ref{sec:appendix1}.

In our study, jet productions in p+p collisions are simulated within a Monte Carlo event generator SHERPA 2.2.11~\cite{Gleisberg:2008ta}, which perform NLO Matrix Element calculations matched to the Parton Shower with several merging schemes. AMEGIC++~\cite{Krauss:2001iv} and Comix~\cite{Gleisberg:2008fv} are inbuilt matrix-element generators of SHERPA, which automatically calculate and integrate tree-level amplitudes, cooperating with their phase-space generator Phasic~\cite{Krauss:2001iv}. OpenLoops~\cite{Cascioli:2011va} programs is interfaced with SHERPA to provide virtual corrections. An method of MC@NLO-type is performed to match a fixed-order next-to-leading order (NLO) calculation with the resummation of the parton shower~\cite{Hoeche:2011fd}. The parton distribution function (PDF) set `NNPDF30$\_$nnlo' is loaded for beams~\cite{Ball:2008by, Ball:2009mk}. For high transverse momentum transfer processes and multi-jet production, LO calculations typically perform worse than NLO ones in describing experimental measurements of the azimuthal angle separation between jets~\cite{CMS:2013lua}. In such cases, SHERPA, by directly incorporating higher-order matrix element calculations, often provides a more accurate description of event kinematics than LO-precision generators~\cite{Sherpa:2019gpd}. Our calculations are performed at the parton level in both p+p and Pb+Pb collisions. 

The spatial structure of $H_1^T$ is depicted in the top-left corner of FIG.~\ref{fig:test}. To gain further insights into the configuration structure of $H_1^T$, we calculate the average azimuthal angle correlations for events with $n_\text{jet} = 2$ and $n_\text{jet} > 2$ in p+p collisions at a center-of-mass energy of $\sqrt{s}$ = 5.02 TeV, considering three logarithmic intervals of $H_1^T$, as presented in TABLE ~\ref{tab1}. The kinetic sets corresponding to this calculation are described in Sec.~\ref{sec:results}. From TABLE ~\ref{tab1}, we observe that the azimuthal angle distance $\langle\Delta\phi_{12}\rangle$ of $n_\text{jet} = 2$ events decreases with increasing $H_1^T$. Additionally, for $n_\text{jet} > 2$ events, both $\langle\Delta\phi_{12}\rangle$ and $\langle\Delta\phi_{13}\rangle$ decrease while $\langle\Delta\phi_{23}\rangle$ increases with increasing $H_1^T$. This indicates that the jet event shape becomes broader as $H_1^T$ increases for both $n_\text{jet} = 2$ and $n_\text{jet} > 2$ events. To present the geometric illustration more vividly, we provide visual schematic diagrams obtained through calculations in FIG.~\ref{config:jet}.

\begin{table}[ht]
  \centering
  \begin{tabular}{|c|c|c|c|c|}
    \hline
    \multirow{2}{*}{$\ln(H_1^T)$} & \text{$n_\text{jet} = 2$ events} & \multicolumn{3}{c|}{\text{$n_\text{jet} > 2$} events} \\ 
    \cline{2-5}
    & $\langle \Delta\phi_{12}\rangle$ & $\langle\Delta\phi_{12}\rangle$ & $\langle\Delta\phi_{13}\rangle$ & $\langle\Delta\phi_{23}\rangle$ \\ \hline
    $\left [-10.0,-7.0 \right ]$ & 3.117 & 2.885 & 2.596 & 0.784 \\ \hline
    $\left [ -7.0,-3.0 \right ]$ & 3.026 & 2.849 & 2.560 & 0.831 \\ \hline
    $\left [ -3.0,0.0 \right ]$ & 2.573 & 2.520 & 2.198 & 1.138 \\ \hline
  \end{tabular}
  \caption{The averaged azimuthal angle correlations are calculated for $n_\text{jet} = 2$ and $n_\text{jet} > 2$ events in p+p collisions at $\sqrt{s}=5.02$ TeV at three $\ln(H_1^T)$ intervals.}
  \label{tab1}
\end{table}

As shown in FIG.~\ref{config:jet}, schematic diagrams illustrating the spatial configurations of $n_\text{jet} = 2$ and $n_\text{jet} = 3$ events across three consecutive $H_1^T$ intervals are plotted based on TABLE ~\ref{tab1}. This schematic employs a scheme in which the direction of the leading jet is the polar axis of the polar coordinates, and in order to better schematize the geometric configuration of the jet events, we rescale the transverse momentum of jets in a dimensionless interval from 0 to 1, preserving the relative proportionality between their transverse momentum magnitudes. These diagrams are visualizations that as $H_1^T$ increases, both $n_\text{jet} = 2$ and $n_\text{jet} = 3$ event shapes become broader.

\section{$\text{FWM}_\text{1st}^\text{T}$ in P\lowercase{b}+P\lowercase{b} collisions}
\label{sec:results}

In this section, we use p+p data generated by the SHERPA event generator as a baseline to explore how the distribution of $H_1^T$ will be modified after undergoing multiple scatterings with nuclear medium partons. In our study, The Linear Boltzmann Transport (LBT) model~\cite{Wang:2013cia, He:2015pra, Cao:2016gvr} is used to simulate the physical process of multiple scattering between jet and medium partons in QGP. Additionally, the information of the bulk partons is provided by (3+1)D CLVisc hydrodynamical model~\cite{Pang:2012he, Pang:2018zzo, Pang:2014ipa, Wu:2021fjf, Jiang:2023fad} with initial conditions simulated from A Multi-Phase Transport (AMPT) model~\cite{Lin:2004en}. Both elastic and inelastic scattering processes are considered for the initial jet shower partons and the thermal recoil partons.

In the LBT model, the $2 \rightarrow 2$ elastic scattering process is simulated by the linear Boltzmann transport equation,
\begin{eqnarray}
	&p_1\cdot\partial f_a(p_1)=-\int\frac{d^3p_2}{(2\pi)^32E_2}\int\frac{d^3p_3}{(2\pi)^32E_3}\int\frac{d^3p_4}{(2\pi)^32E_4} \nonumber \\
	&\frac{1}{2}\sum _{b(c,d)}[f_a(p_1)f_b(p_2)-f_c(p_3)f_d(p_4)]|M_{ab\rightarrow cd}|^2 \nonumber \\
	&\times S_2(s,t,u)(2\pi)^4\delta^4(p_1+p_2-p_3-p_4)
	\label{lbt}
\end{eqnarray}
 
The inelastic scattering, what is called the medium-induced gluon radiation in the LBT model, is described by the higher twist formalism~\cite{Guo:2000nz, Zhang:2003wk, Zhang:2003yn, Majumder:2009ge},
\begin{equation}
	\frac{dN_g}{dxdk_\perp^2 dt}=\frac{2\alpha_sC_AP(x)\hat{q}}{\pi k_\perp^4}\left(\frac{k_\perp^2}{k_\perp^2+x^2M^2}\right)^2\sin^2\left(\frac{t-t_i}{2\tau_f}\right)
  \label{ht_eq}
\end{equation}

We refer to ~\cite{Wang:2013cia, He:2015pra, Cao:2016gvr} for the entensive discussions on the interpretation of various symbols and implementations of the elastic and inelastic scattering of fast partons with the medium in Eq.~(\ref{lbt}) and Eq.~(\ref{ht_eq}).

\begin{figure*}[ht]
  \centering
  \vspace{0.in}
  \hbox{
      \includegraphics[width=17cm,height=8.5cm]{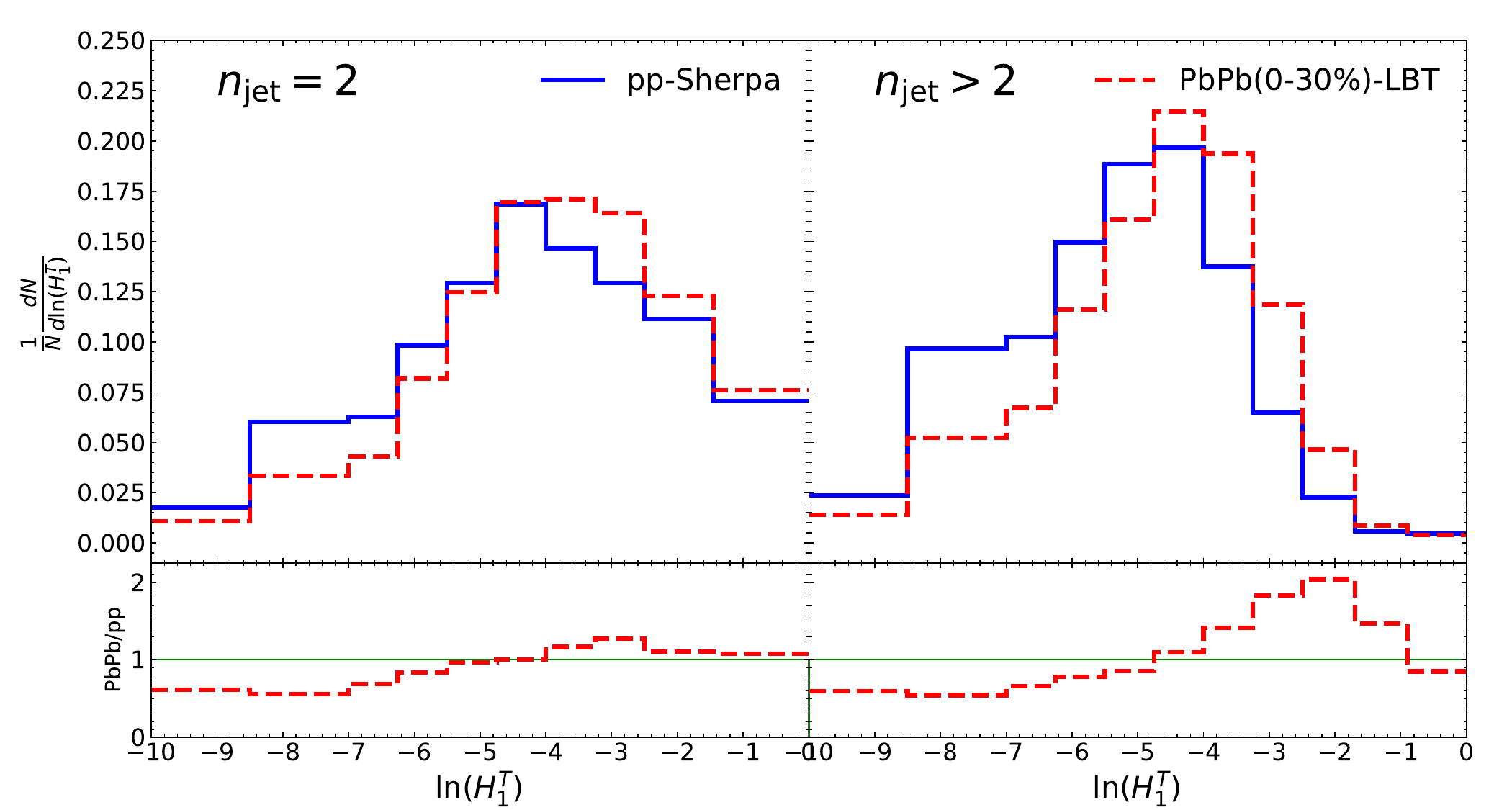}
      }
  \caption{The event normalized $H_1^T$ distributions for $n_\text{jet} = 2$ (left panel) and $n_\text{jet} > 2$ (right panel) events in p+p and Pb+Pb collisions at $\sqrt{s}=5.02$ TeV.}
  \label{jetbin}
\end{figure*}

By generating p+p events with Sherpa + OpenLoops program packages and simulating the production of jets in Pb+Pb collisions at $\sqrt{s_{NN}} = 5.02$ TeV with LBT model, we present our results and analyze them in this section. The anti-$k_{\perp}$ algorithm is used to reconstruct jets with a cone size of R = 0.4 using the FastJet package~\cite{Cacciari:2011ma}. The jet events are required to have at least two in our study. The kinematic sets are as follows: the transverse momentum of leading jet $p_T^\text{leading} > 110$ GeV/c, the threshold $p_T^\text{jet} > 30$ GeV/c and rapidity $|\eta| \le$ 2.5 for all jets.

We start the investigation by studying the nuclear modifications of the $H_1^T$ distributions for events with a fixed number of jets. The jet events are classified into two types, events with $n_\text{jet} = 2$ and $n_\text{jet} > 2$.

Let us first focus on the event normalized spectrum of the $H_1^T$ for the $n_\text{jet} = 2$ events and its nuclear modification, as shown in the left panel of FIG.~\ref{jetbin}. The left panel of FIG.~\ref{jetbin} shows the results of the p+p (black line) and Pb+Pb (red line) collisions at the $\sqrt{s} = 5.02$ TeV, and the ratio of these two, also known as the nuclear modification factor, is shown in the bottom panel. The distribution is apparently suppressed and enhanced separately in the small and large regions, taking approximately $\ln(H_1^T) = {\color{red}-4.5}$ as a split. This medium modification means that the proportion of back-to-back jets, which are events with narrow shapes, is reduced, and therefore the proportion of $n_\text{jet} = 2$ events is enhanced in regions characterized by large $H_1^T$, where the event shapes are broader. The jet quenching effect indicates that the jet primarily loses energy through medium-induced gluon radiation. This leads to a more diffusely distributed total energy after passing through the medium, resulting in the broadening of $n_\text{jet} = 2$ events.

\begin{table*}[htb]
  \centering
  \scalebox{1.35}{
  \begin{tabular}{|c|c|c|c|c|c|c|c|}
    \hline
    \multicolumn{2}{|c|}{\text{$n_\text{jet} = 2$ events}} & \multicolumn{6}{c|}{\text{$n_\text{jet} > 2$} events} \\ \hline
    p+p & Pb+Pb & \multicolumn{3}{c|}{\text{p+p}} & \multicolumn{3}{c|}{\text{Pb+Pb}} \\ \hline
    $\langle \Delta\phi_{12}\rangle$ & $\langle\Delta\phi_{12}\rangle$ & $\langle\Delta\phi_{12}\rangle$ & $\langle\Delta\phi_{13}\rangle$ & $\langle\Delta\phi_{23}\rangle$ & $\langle\Delta\phi_{12}\rangle$ & $\langle\Delta\phi_{13}\rangle$ & $\langle\Delta\phi_{23}\rangle$ \\ \hline
    2.946 & 2.922 & 2.829 & 2.539 & 0.847 & 2.816 & 2.465 & 0.934 \\ \hline
  \end{tabular}
  }
  \caption{The averaged azimuthal angle correlations are calculated for $n_\text{jet} = 2$ and $n_\text{jet} > 2$ events in p+p and Pb+Pb collisions at $\sqrt{s}=5.02$ TeV, across the entire $\ln(H_1^T)$ statistical region.}
  \label{tab2}
\end{table*}

\begin{figure}[!htb]
  \centering
  \vspace{0.in}
  \includegraphics[width=8cm,height=7.5cm]{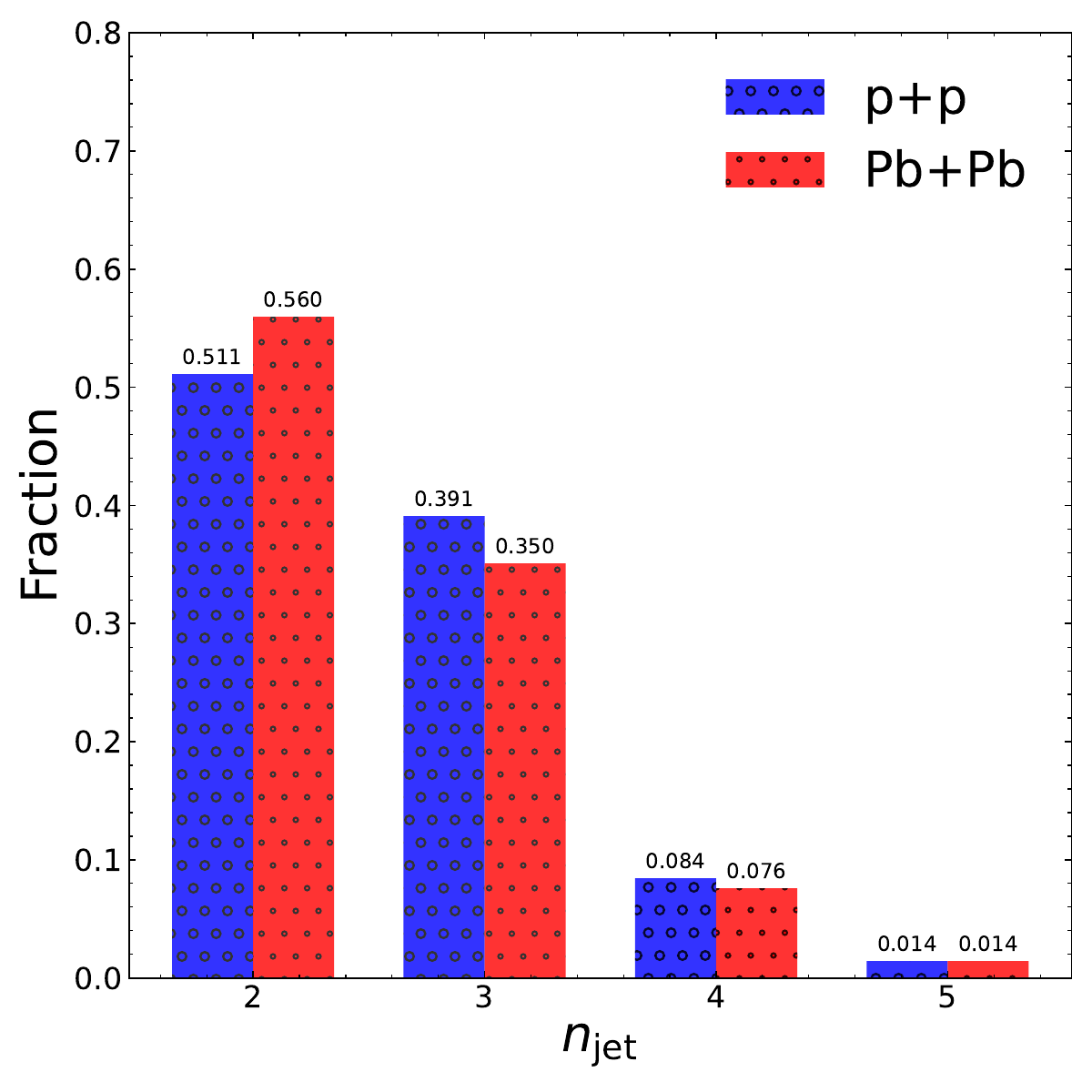}
  \caption{Histograms of the fraction of different numbers of jets in all jet events in p+p and Pb+Pb collisions at $\sqrt{s}$ = 5.02 TeV.}
  \label{jn}
\end{figure}

Subsequent to the preceding paragraph, the event normalized $H_1^T$ spectrum of $n_\text{jet} > 2$ events in p+p collisions (black line) and Pb+Pb collisions (red line) and their nuclear modifications are shown in the right panel of FIG.~\ref{jetbin}. The distributions exhibit an significant enhancement in the bins where $\ln(H_1^T) \in$ [-4.8, -0.9] by the medium, with suppressions observed in the other regions. The reason for the medium modification in the bins $\ln(H_1^T) \in [-10, -0.9]$ is the same as for $n_{\text{jet}} = 2$, as mentioned above, which is due to the spread of momenta in the medium. There is another thing that should be considered, which is the jet number reduction effect~\cite{Apolinario:2012cg} when we focus on $n_\text{jet} > 2$ events (see FIG.~\ref{jn}). It leads to the proportion of $n_\text{jet} > 2$ events reduce significantly in the large region where $\ln(H_1^T) \in$ [-0.9, 0].

To more intuitively reflect the medium modification of the event shape, we present the average azimuthal angle distance between jets for events with $n_\text{jet} = 2$ and $n_\text{jet} > 2$ over the entire $\ln(H_1^T)$ statistical region in TABLE ~\ref{tab2}. It should be noted that we only consider the top three jets ranked by transverse momentum for $n_\text{jet} > 2$ events. As seen in TABLE ~\ref{tab2}, the shape of $n_\text{jet} = 2$ events broadens, with the average azimuthal angle distance $\langle \Delta\phi_{12} \rangle$ decreasing due to medium modification. For $n_\text{jet} > 2$ events, both $\langle \Delta\phi_{12} \rangle$ and $\langle \Delta\phi_{13} \rangle$ decrease, while $\langle \Delta\phi_{23} \rangle$ increases, indicating that the event shape is also broadened by medium modification.

The histograms showing fractions of different numbers of jets in total jet events for p+p and Pb+Pb collisions at $\sqrt{s} = 5.02$ TeV are presented in FIG.~\ref{jn}. In Pb+Pb collisions, the change in the number of jets as a fraction of total jet events is significant compared to p+p collisions. The fraction of events with $n_\text{jet} = 2$, increases from 0.51 in p+p collisions to 0.56 in Pb+Pb collisions, and correspondingly, the total fraction of other events, i.e., $n_\text{jet} > 2$ events, decreases from 0.49 to 0.44. The decrease in the number of events is known as the jet number reduction effect. The reason for this is that some of the sub-leading jets experience energy loss in the medium, causing their transverse momentum magnitude to fall below our observational threshold of 30 GeV/c. As a result, the fraction of observed $n_\text{jet} = 2$ events increases in Pb+Pb collisions compared to p+p collisions.

\begin{figure}
  \vspace{0.in}
  \includegraphics[width=8.0cm,height=8.0cm]{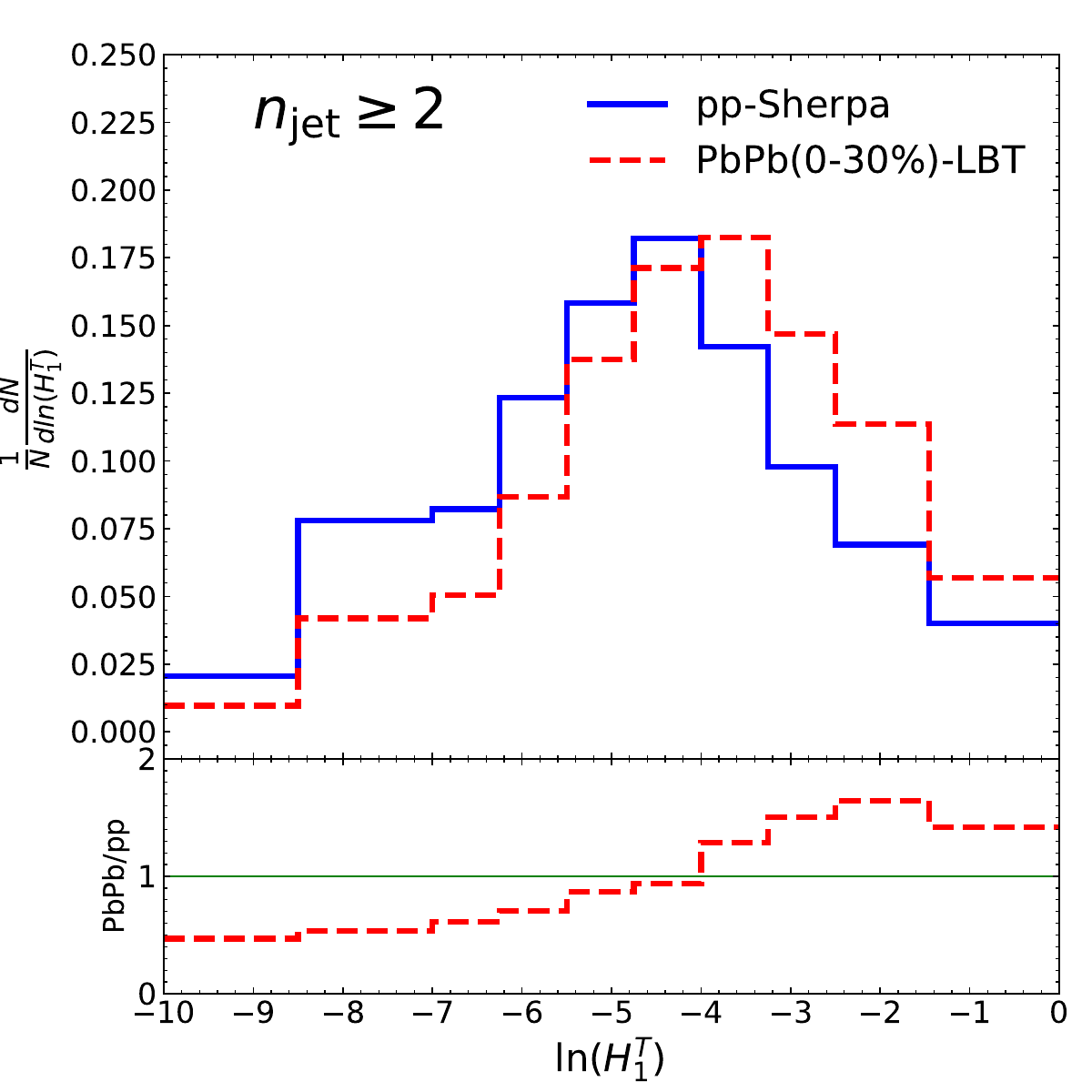} \\
  \caption{Top: event normalized $H_1^T$ distribution of $n_\text{jet} \ge 2$ events in p+p and Pb+Pb collisions at $\sqrt{s}=5.02$~TeV;
  Bottom: medium modification factor of event normalized $H_1^T$ distribution of $n_\text{jet} \ge 2$ events.}
  \label{H1_tot}
\end{figure}

\begin{figure}
  \centering
  \vspace{0.in}
  \includegraphics[width=8.0cm,height=8.0cm]{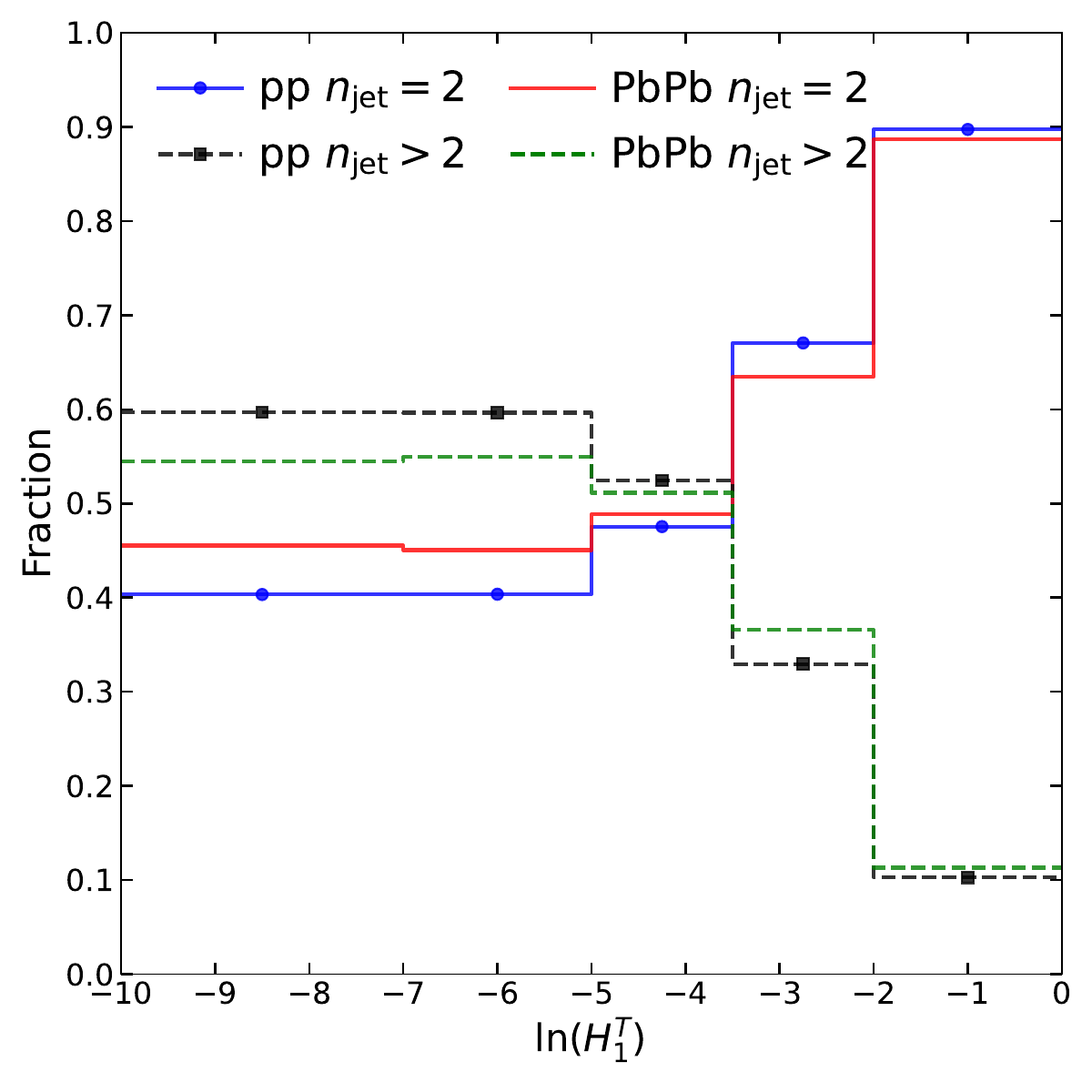} \\
  \vspace{-0.1in}
  \caption{The relative contribution fractions of $n_\text{jet} = 2$ and $n_\text{jet} > 2$ events from the distribution of $H_1^T$ normalized by the total number of events in p+p and Pb+Pb collisions at $\sqrt{s} =$ 5.02 TeV.}
  \label{frac}
\end{figure}

\begin{figure*}[hbt]
  \centering
  \includegraphics[width=17.cm,height=7.5cm]{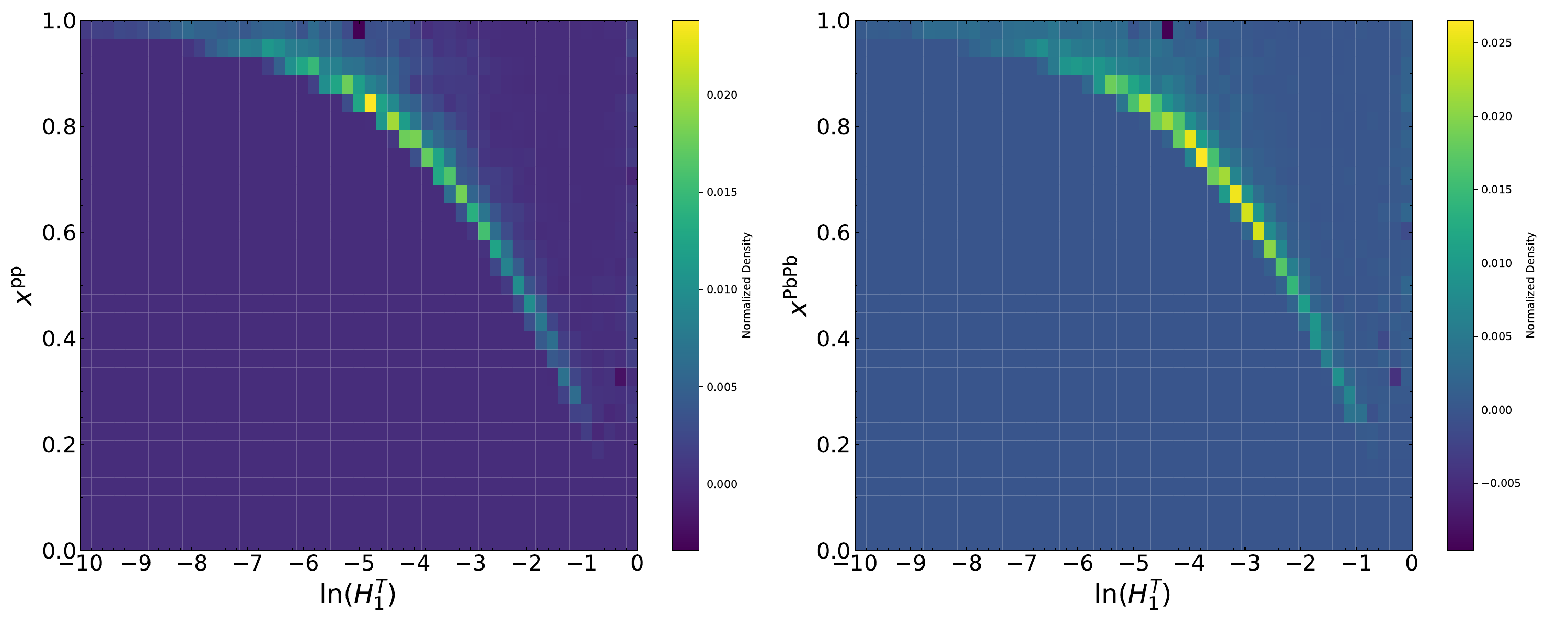}
  \caption{Distribution correlation between $H_1^T$ and x for $n_\text{jet} = 2$ event in p+p (left) and Pb+Pb (right) system at $\sqrt{s} = 5.02$ TeV.}
  \label{HX}
\end{figure*}

Now we consider the $H_1^T$ distribution for the combination of $n_\text{jet} = 2$ and $n_\text{jet} > 2$ events in proportion, as shown in FIG.~\ref{H1_tot}. The nuclear modification factor exhibits a similar pattern to that of $n_\text{jet} = 2$ events, characterized by suppression at low $H_1^T$ values and enhancement at high $H_1^T$ values. To understand why the nuclear medium modification for total ($n_\text{jet} \geq 2$) events is as shown in FIG.~\ref{H1_tot}, we draw FIG.~\ref{frac}, which illustrates the relative contributions of $n_\text{jet} = 2$ and $n_\text{jet} > 2$ events to $H_1^T$ in both p+p and Pb+Pb collisions at $\sqrt{s} = 5.02$ TeV. In the low $H_1^T$ region, $n_\text{jet} > 2$ events constitute a slightly larger proportion than $n_\text{jet} = 2$ events. As $H_1^T$ increases, the percentage of $n_\text{jet} = 2$ ($n_\text{jet} > 2$) events gradually rises from approximately 40\% (60\%) in low the $H_1^T$ region to around 90\% (10\%) in the high $H_1^T$ region in p+p collisions. Similarly, the corresponding percentages transition from approximately 46\% (54\%) to around 89\% (11\%) in Pb+Pb collisions. The transition in the relative contributions between $n_\text{jet} = 2$ and $n_\text{jet} > 2$ events occurs at around $\ln(H_1^T) = -4$. The nuclear modification remains suppressed in the low $H_1^T$ region, where both $n_\text{jet} = 2$ and $n_\text{jet} > 2$ events have similar fractions. In contrast, an enhancement effect is observed in the high $H_1^T$ region, consistent with the behavior of $n_\text{jet} = 2$ events, which dominate in this region.

Although the focus of this work is on the study of jet event shapes described by $H_1^T$, rather than the transverse momentum or energy asymmetry, the latter nonetheless defines the former. We also discuss the medium modification of the transverse momentum asymmetry for $n_\text{jet} = 2$ events in Appendix~\ref{sec:appendix2}.

\section{summary}
\label{sec:sum}

We utilize the Monte Carlo event generator SHERPA to simulate jet production in p+p collisions. To incorporate virtual corrections, the OpenLoops programs are seamlessly integrated into SHERPA environment. Additionally, our study employs LBT model to replicate the intricate phenomenon of multiple scatterings between jets and the medium.

We investigate $H_1^T$, which is the lower-order member of the set of FWMs related to QCD geometric patterns, and apply it to the field of jet quenching physics. This application allows us to examine how the medium modifies jet event shape of varying multiplicities ($n_\text{jet}$). The results indicate that $n_\text{jet} = 2$ events undergo a pronounced broadening modification due to the influence of nuclear matter. Meanwhile, the event shape for $n_\text{jet} > 2$ becomes broader due to medium modification, except in the highest $\ln(H_1^T)$ bin where the jet number reduction effect is significant. Additionally, we investigate the underlying reasons for the medium modification of $H_1^T$ spectra for total jet events, without distinguishing between different multiplicities. This is achieved by analyzing the relative fraction of $H_1^T$ between the two types of events.

\bigskip

{\bf Acknowledgments:}  
We extend our sincere gratitude to Jin-Wen Kang, Wei Dai, Shi-Yong Chen, Meng-Quan Yang, Sa Wang, and Shan-Liang Zhang for their invaluable discussions. We would also like to express our appreciation to the open-source software packages Sherpa~\cite{Gleisberg:2008ta}, FastJet~\cite{Cacciari:2011ma}, HepMC2~\cite{Dobbs:2001ck} and ROOT~\cite{Brun:1997pa}. This work has been supported by the Guangdong Major Project of Basic and Applied Basic Research under Grant No. 2020B030103008, as well as the Natural Science Foundation of China with Project Nos. 11935007 and 12035007.

\bigskip

\appendix
\section{Formal simplification of the $H_1^T$ definition for $n_\text{jet} = 3$}
\label{sec:appendix1}

The definition of $H_1^T$ is complex when $n_\text{jet} > 2$ and is not discussed in detail in the main text. Here, we discuss the case of $n_\text{jet} = 3$ as an example.

\begin{equation}
  \begin{aligned}
        H_1^{T} & = \sum_{i,j=1}^{3} \; \frac{p_{T i}p_{T j}}  {\left(\sum_{k=1}^{3} p_{T k}\right)^2} \; \cos \Delta\phi_{ij}\\
    \label{eq:njet3}
  \end{aligned}
\end{equation}

Following Eq.~\eqref{eq:fwm_toy}, we denote $p_{T 2} = y\cdot p_{T 1} $ and $p_{T 3} = z\cdot p_{T 1} $ which simplify the functional form of Eq.~\eqref{eq:njet3}. We get:

\begin{equation}
  \begin{aligned}
        H_1^{T} & =  \frac{1 + y^2 + z^2 + 2y\cos\phi_{12} + 2z\cos\phi_{13} + 2yz\cos\phi_{23}}  {1 + y^2 + z^2 + 2y + 2z +2yz}
    \label{eq:njet3_sim}
  \end{aligned}
\end{equation}

The above simplification is intended to inform readers of a form similar to Eq.~\eqref{eq:fwm_toy}; however, due to the increase in the number of variables, we have not conducted an in-depth analysis. Nonetheless, as discussed above, the physical interpretation of $H_1^T$ in events with $n_\text{jet} \geq 2$ is that the event shape broadens as $H_1^T$ increases.

\section{Correlations between $H_1^T$ and x}
\label{sec:appendix2}

To further explore the factors responsible for the medium modification of the $H_1^T$ distribution, we select $n_\text{jet} = 2$ events for correlation studies. The transverse momentum asymmetry is defined as $x = p_{T 2}/p_{T 1}$. We plot the distribution correlation between $H_1^T$ and x for $n_\text{jet} = 2$ events in both p+p and Pb+Pb systems at $\sqrt{s}$ = 5.02 TeV in FIG.~\ref{HX}. Remarkably pronounced medium modification phenomena can be discerned from this result. It can be inferred that the medium modification of $H_1^T$ exhibits moderate sensitivity to the transverse momentum imbalance between jets, which accentuates the transverse momentum asymmetry.

\bigskip

\bibliographystyle{apsrev4-1}
\bibliography{FWM_v8}

\end{document}